\documentclass[11pt, reqno]{{amsart}}
\usepackage{amsfonts}
\usepackage{amssymb}

\def\Box{\vcenter{\vbox{\hrule\hbox{\vrule
     \vbox to 8.8pt{\hbox to 10pt{}\vfill}\vrule}\hrule}}}

\numberwithin{equation}{section}

\begin{document}

\title[A Class of Nonbinary Codes and Their Weight Distribution]
 {A Class of Nonbinary Codes and Their Weight Distribution}

\author[Xiangyong Zeng, Nian Li, and Lei Hu ]
{Xiangyong Zeng, Nian Li, and Lei Hu }


\address{Xiangyong Zeng and Nian Li are with the Faculty of Mathematics and Computer Science,
Hubei University, Wuhan, China. Email: xzeng@hubu.edu.cn}

 \address{Lei Hu is with the State Key Laboratory of
Information Security, Graduate University of Chinese Academy of
Sciences, Beijing, China. Email: hu@is.ac.cn}


\date{}

\begin{abstract}
In this paper, for an even integer $n\geq 4$ and any positive
integer $k$ with ${\rm gcd}(n/2,k)={\rm gcd}(n/2-k,2k)=d$ being odd,
a class of $p$-ary codes $\mathcal{C}^k$ is defined and their weight
distribution is completely determined, where $p$ is an odd prime. As
an application, a class of nonbinary sequence families is
constructed from these codes, and the correlation distribution is
also determined.
\end{abstract}
\maketitle

\section{Introduction}\label{sec-intro}

Nonlinear functions have important applications in coding theory and
cryptography \cite{MS, GG}. Linear codes constructed from functions
with high nonlinearity \cite{R, KSW,CD, DY} can be good and have
useful applications in communications \cite{K,SP, KM, SOSL} or
cryptography \cite{DH,CDY,CDN, YCD}. For a code, its weight
distribution is important to study its structure and to provide
information on the probability of undetected error when the code is
used for error detection.

\vspace{1mm}

Throughout this paper, let $\mathbb{F}_{q}$ be the finite field with
$q=p^n$ elements for a prime $p$ and a positive integer $n$, and let
$\mathbb{F}^*_{q}$ be the multiplicative group of $\mathbb{F}_{q}$.
For an even integer $n\geq 4$, let $\mathcal{C}^{k}$ denote the
$[p^n-1,5n/2]$ cyclic code given by
$$\begin{array}{c}\mathcal{C}^{k}={\Big\{}c(\gamma,\delta,\epsilon)=\big(\Pi_{\gamma,\delta}(x)+Tr^n_1(\epsilon
x)\big)_{x\in \mathbb{F}_{p^n}^*}\mid\gamma \in
\mathbb{F}_{p^{n/2}},\, \delta, \,\epsilon\in \mathbb{F}_{p^n}
{\Big\}}\end{array}$$ constructed from the function
\begin{equation}\begin{array}{c}\Pi_{\gamma,\delta}(x)=Tr_1^{n/2}(\gamma
x^{p^{n/2}+1}) +Tr_1^n(\delta x^{p^k+1}),\end{array}\end{equation}
where $1\leq k<n$ with $k\neq n/2$, and for a positive integer $l$,
$Tr^l_1(\cdot)$ is the trace function from $\mathbb{F}_{p^l}$ to
$\mathbb{F}_{p}$.

\vspace{1mm}

Several classes of binary codes $\mathcal{C}^{k}$ have been
extensively studied for some values of the parameter $k$. The binary
code $\mathcal{C}^{n/2\pm 1}$ is exactly the Kasami code in Theorem
14 of \cite{K}. By choosing cyclicly inequivalent codewords from the
Kasami code, the large set of binary Kasami sequences was obtained
\cite{SP}. The minimum distance bound of $\mathcal{C}^{1}$ was
established by evaluating the exponential sums
$\begin{array}{c}\sum\limits_{x\in
\mathbb{F}_{2^n}}(-1)^{\Pi_{\gamma,\delta}(x)+Tr^n_1(\epsilon
x)}\end{array}$ in \cite{MK} and \cite{L}, and the weight
distribution and then the minimum distance were completely
determined in \cite{V}. For even $n/2$, the binary code
$\mathcal{C}^{1}$ has the same weight distribution as the Kasami
code \cite{K,V}. Furthermore, for any $k$ with ${\rm gcd}(k,n)=2$ if
$n/2$ is odd or ${\rm gcd}(k,n)=1$ if $n/2$ is even, the weight
distribution of binary codes $\mathcal{C}^{k}$ was also determined,
and these codes were used to construct families of generalized
Kasami sequences, which have the same correlation distribution and
family size as the large set of Kasami sequences \cite{ZLH}.

\vspace{1mm}

The purpose of this paper is to study the weight distribution of the
code $\mathcal{C}^{k}$ in the nonbinary case, namely we assume $p$
is odd, for a wide range of $k$ that satisfies
\begin{equation}{\rm gcd}(n/2,k)={\rm gcd}(n/2-k,2k)=d\,\, {\rm being\,\, odd}.\end{equation}
Applying the techniques developed in \cite{B}, we describe some
properties of the roots to the equation
$$\delta^{p^{n-k}}y^{p^{n/2-k}+1}+\gamma y+\delta=0$$
with $\gamma\delta\not=0$. Based on these properties and the theory
of quadratic theory over finite fields of odd characteristic, we
completely determine the weight distribution. As an application,
these codes are also used to construct a class of nonbinary sequence
families. \vspace{1mm}

The remainder of this paper is organized as follows. Section 2 gives
some preliminaries and the main result. Section 3 considers the rank
distribution of a class of quadratic forms. Section 4 determines the
weight distribution of the nonbinary codes, and these codes are used
to construct a class of nonbinary sequence families with low
correlation in Section 5. Section 6 concludes the study.

\section{Preliminaries and Main Result}

For positive integers $n$ and $l$ with $l$ dividing $n$, the trace
function $Tr^n_l(\cdot)$ from $\mathbb{F}_{p^n}$ to
$\mathbb{F}_{p^l}$ is defined by
$$\begin{array}{c}
Tr^n_l(x)=\sum\limits^{n/l-1}_{i=0}x^{p^{li}}, \;\;\;  x \in
\mathbb{F}_{p^n}.\end{array}$$ For the properties of the trace
function, please see \cite{LN}.

Let $q=p^n$. The field $\mathbb{F}_{q}$ is an $n$-dimensional vector
space over $\mathbb{F}_{p}$. For any given basis $\{\alpha_1,
\alpha_2,\cdots,\alpha_n\}$ of  $\mathbb{F}_{q}$ over
$\mathbb{F}_{p}$, each element $x\in \mathbb{F}_{q}$ can be uniquely
represented as $x=x_1\alpha_1+x_2\alpha_2+\cdots +x_n\alpha_n$ with
$x_i\in \mathbb{F}_p$ for $1\leq i\leq n$. Under this
representation, the field $\mathbb{F}_{q}$ is identical to the
$\mathbb{F}_p$-vector space $\mathbb{F}_p^n$. A function $f(x)$ on
$\mathbb{F}_{p^n}$ is a {\it quadratic form} if it can be written as
a homogeneous polynomial of degree 2 on $\mathbb{F}^n_p$, namely of
the form
$$\begin{array}{c}f(x_1,\cdots,x_n)=\sum\limits_{1\leq i\leq j\leq n}a_{ij}x_ix_j\end{array}$$
where $a_{ij}\in \mathbb{F}_p$. The {\it rank of the quadratic form}
$f(x)$ is defined as the codimension of the $\mathbb{F}_p$-vector
space
\begin{equation}\begin{array}{c}V_f={\Big\{}z\in \mathbb{F}_{p^n}\,|\,f(x+z)=f(x)\,\,{\rm for}\,\,{\rm all}\,\,x\in \mathbb{F}_{p^n}{\Big\}},\end{array}\end{equation}  denoted by ${\rm rank}(f)$.
Then $|V_f|=p^{n-{\rm rank}(f)}$.

\vspace{1mm}

For the quadratic form $f(x)$, there exists a symmetric matrix $A$
such that $$f(x)=X^{{\rm T}}AX,$$
 where $X^{{\rm
T}}=(x_{1}, x_{2},\cdots, x_{n})\in\mathbb{F}_{p}^n$ denotes the
transpose of a column vector $X$. The {\it determinant ${\rm
det}(f)$} of $f(x)$ is defined to be the determinant of $A$, and
$f(x)$ is {\it nondegenerate} if ${\rm det}(f)\neq 0$. By Theorem
6.21 of \cite{LN}, there exists a nonsingular matrix $B$ such that
$B^{{\rm T}}AB$ is a diagonal matrix. Making a nonsingular linear
substitution $X=BY$ with $Y^{{\rm T}}=(y_{1}, y_{2},\cdots, y_{n})$,
one has
\begin{equation}
f(x)=Y^{{\rm T}}B^{{\rm T}}ABY= \sum\limits_{i=1}^{n}a_{i}y_i^2
\end{equation}
for $a_1,a_2,\cdots,a_n\in \mathbb{F}_{p}$. Notice that a degenerate
quadratic form $f(x)$ over $\mathbb{F}_p^n$ is possibly
nondegenerate over $\mathbb{F}_p^t$ ($t<n$) after a nonsingular
substitution.

The {\it quadratic character} of $\mathbb{F}_q$ is defined by
$$
\eta(x)=\left \{
 \begin{array}{cl}
1,&{\rm if\,\,} x {\rm\,\,is \,\,a\,\, square \,\,element\,\,in}\,\, \mathbb{F}^*_q,\\
-1, &{\rm if\,\,} x {\rm\,\,is \,\,a\,\,\,nonsquare \,\,element\,\,in}\,\, \mathbb{F}^*_q,\\
0, &{\rm if\,\,}x=0.
\end{array}
 \right.
$$

The following two lemmas about quadratic form will be frequently
used to prove the results of this paper.

\vspace{1mm}

{\it Lemma 1 ( Theorems 6.26 and 6.27 of \cite{LN}): } For odd $q$,
let $f$ be a nondegenerate quadratic form over $\mathbb{F}_q$ in $l$
indeterminates, and a function $\upsilon(x)$ over $\mathbb{F}_q$
defined by
$$
\upsilon(x)=\left \{
 \begin{array}{cl}
-1,&{\rm if\,\,} x \in \mathbb{F}^*_q,\\
q-1, &{\rm otherwise}.
\end{array}
 \right.
$$ Then for $\rho\in \mathbb{F}_q$ the number of solutions to the
equation $f(x_1,\cdots,x_n)=\rho$ is $$
q^{l-1}+q^{\frac{l-1}{2}}\eta\left((-1)^{\frac{l-1}{2}}\rho\cdot{\rm
det}(f)\right)$$ for odd $l$, and
$$q^{l-1}+\upsilon(\rho)q^{\frac{l-2}{2}}\eta\left((-1)^{\frac{l}{2}}
{\rm det}(f)\right)$$ for even $l$.

 {\it Lemma 2 (Theorem 5.15 of \cite{LN}):} Let $\omega$ be a complex
 primitive $p$-th root of unity. Then
$$
\sum\limits_{k=1}^{p-1}\eta(k)\omega^{k}=\sqrt{(-1)^{\frac{p-1}{2}}p}.$$

\vspace{1mm}

A {\it $p$-ary $[m,l]$ linear code} $\mathcal{C}$ is a linear
subspace of $\mathbb{F}_{p}^m$ with dimension $l$. The {\it Hamming
weight} of a codeword $c_1c_2\cdots c_m$ of $\mathcal{C}$ is the
number of nonzero $c_i$ for $1\leq i\leq m$.

\vspace{1mm}

Let $\mathcal{F}$ be a family of $M$ $p$-ary sequences of period
$p^n-1$ given by
$$
\mathcal{F}={\Big\{}\{s_i(t)\}_{t=0}^{p^n-2}\,\,|\,0\leq i\leq
M-1{\Big\}}.
$$
The {\it periodic correlation function} of the sequences
$\{s_i(t)\}$ and $\{s_j(t)\}$ in $\mathcal{F}$ is
$$
 C_{i,j}(\tau)=\sum\limits^{p^n-2}_{t=0}\omega^{s_i(t)-s_j(t+\tau)}
$$ where $0\leq \tau\leq p^n-2$. The two sequences $\{s_i(t)\}$ and $\{s_j(t)\}$ in
$\mathcal{F}$ is {\it cyclicly inequivalent} if
$|C_{i,j}(\tau)|<p^n-1$ for any $\tau$. The {\it maximum magnitude}
$C_{\rm max}$ of the correlation values is
$$
C_{\rm max}={\rm max} \{|C_{i,j}(\tau)|: i\neq j\,\,{\rm or}\,\,
\tau\neq 0\}.
$$

From now on, we always assume that the prime $p$ is odd, and
$n=2m\geq 4$.

\vspace{1mm}

Since $Tr_{1}^{n}(\delta x^{p^k+1})=Tr_{1}^{n}(\delta^{p^{n-k}}
x^{p^{n-k}+1})$ and $\delta^{p^{n-k}}$ runs through
$\mathbb{F}_{p^n}$ as $\delta$ runs through $\mathbb{F}_{p^n}$,
without loss of generality, the integer $k$ in the definition of
code $\mathcal{C}^k$ is assumed to satisfy $1\leq k<n/2$. For an odd
integer $t$ relatively prime to $m$, the integer $k=m-t$ satisfies
Equality (1.2), and $d=1$. In particular, for $t=1$, ${\rm
gcd}(m,k)={\rm gcd}(m-k,2k)=1$. The parameter $k=m-1$ corresponds to
the binary Kasami code \cite{K}, and for this reason, we call these
$p$-ary $[p^n-1,5m]$ linear codes $\mathcal{C}^k$ with $k$
satisfying Equality (1.2) {\it the nonbinary Kasami codes}.

The main result of this paper is stated as the following theorem.

{\it Theorem 1:} For an even integer $n=2m\geq 4$ and any positive
integer $k$ satisfying Equality (1.2), the weight distribution of
the nonbinary Kasami codes $\mathcal{C}^k$ is given as Table 1.

\vspace{1mm}

This theorem will be proven by the techniques developed in the next
two sections.

\begin{table}\caption{Weight distribution of the nonbinary Kasami codes $\mathcal{C}^k$}
\begin{center}
\begin{tabular}{|c|c|}
\hline weight &  Frequency \\
\hline
 $0$& $1$\\
\hline $(p-1)p^{n-1}$& $(p^n-1)(1+p^{m+n-d}-p^{m+n-2d}+p^{m+n-2d-1}+p^{m+n-3d}-p^{n-2d})$\\
\hline $(p-1)(p^{n-1}-p^{\frac{n-2}{2}})$& $p^d(p^m+1)(p^n-1)(p^{n-1}+(p-1)p^{\frac{n-2}{2}})/\left(2(p^d+1)\right)$\\
\hline $(p-1)(p^{n-1}+p^{\frac{n-2}{2}})$& $(p^{n+d}-2p^n+p^d)(p^m-1)(p^{n-1}-(p-1)p^{\frac{n-2}{2}})/\left(2(p^d-1)\right)$\\
\hline $(p-1)p^{n-1}+p^{\frac{n-2}{2}}$& $p^d(p^m+1)(p^n-1)(p-1)(p^{n-1}-p^{\frac{n-2}{2}})/\left(2(p^d+1)\right)$\\
\hline $(p-1)p^{n-1}-p^{\frac{n-2}{2}}$& $(p^{n+d}-2p^n+p^d)(p^m-1)(p-1)(p^{n-1}+p^{\frac{n-2}{2}})/\left(2(p^d-1)\right)$\\
\hline $(p-1)p^{n-1}-p^{\frac{n+d-1}{2}}$& $p^{m-d}(p^n-1)(p-1)(p^{n-d-1}+p^{\frac{n-d-1}{2}})/2$\\
\hline $(p-1)p^{n-1}+p^{\frac{n+d-1}{2}}$& $p^{m-d}(p^n-1)(p-1)(p^{n-d-1}-p^{\frac{n-d-1}{2}})/2$\\
\hline $ (p-1)(p^{n-1}+p^{\frac{n+2d-2}{2}}) $& $\,\,(p^{m-d}-1)(p^n-1)\left(p^{n-2d-1}-(p-1)p^{\frac{n-2d-2}{2}}\right)/(p^{2d}-1)$\\
\hline
$(p-1)p^{n-1}-p^{\frac{n+2d-2}{2}}$& $(p^{m-d}-1)(p^n-1)(p-1)(p^{n-2d-1}+p^{\frac{n-2d-2}{2}})/(p^{2d}-1)\,\,$\\
 \hline
\end{tabular}
\end{center}
\end{table}

\vspace{2mm}

\section{Rank Distribution of Quadratic Form $\Pi_{\gamma,\delta}(x)$}

This section investigates the rank distribution of the quadratic
form $ \Pi_{\gamma,\delta}(x)$ defined by Equality (1.1) for either
$\gamma\neq 0$ or $\delta\neq 0$.

\vspace{1mm}

The possible rank values of $ \Pi_{\gamma,\delta}(x)$ have a close
relationship with the properties of the roots of the polynomial
$g_{\delta,\gamma}(y)$ in the following Proposition 1, which can be
proven by applying the following lemma introduced by Bluher
\cite{B}.

{\it Lemma 3 (Theorems 5.4 and 5.6 of \cite{B}):} Let
$h_c(x)=x^{p^s+1}-cx+c$, $c\in \mathbb{F}_{p^l}^*$. Then $h_c(x)=0$
has either $0$, $1$, $2$, or $p^{{\rm gcd}(s,l)}+1$ roots in
$\mathbb{F}_{p^l}^*$. Further, let $N_1$ denote the number of $c\in
\mathbb{F}_{p^l}^*$ such that $h_c(x)=0$ has exactly one solution in
$\mathbb{F}_{p^l}^*$, then $N_1=p^{l-{\rm gcd}(s,l)}$ and if $x_0
\in \mathbb{F}_{p^l}^*$ is the unique solution of the equation, then
$(x_0-1)^{\frac{p^l-1}{p^{{\rm gcd}(s,l)}-1}}=1$.

\vspace{1mm}

{\it Proposition 1:} Let
$g_{\delta,\gamma}(y)=\delta^{p^{n-k}}y^{p^{m-k}+1}+\gamma y+\delta$
with $\gamma\delta\not=0$, and $d$ be defined as in Equality (1.2).
Then

\vspace{1mm}

(1) The equation $g_{\delta,\gamma}(y)=0$ has either $0$, $1$, $2$,
or $p^d+1$ roots in $\mathbb{F}_{p^n}$;

(2) If $y_1$ and $y_2$ are two different solutions of
$g_{\delta,\gamma}(y)=0$, then $(y_1y_2)^\frac{p^n-1}{p^d-1}=1$;

(3) If $g_{\delta,\gamma}(y)=0$ has exactly one solution $y_0\in
\mathbb{F}_{p^n} $,  then $y_0^\frac{p^n-1}{p^d-1}=1$.

 \vspace{1mm}

{\it Proof:} (1) Let $y=-\frac{\delta}{\gamma}x$ and
$c=\frac{\gamma^{p^{m-k}+1}}{\delta^{p^{m-k}(p^m+1)}}$.  The
equation $g_{\delta,\gamma}(y)=0$ is equivalent to
\begin{equation} x^{p^{m-k}+1}-cx+c=0.
\end{equation}
Since $c\in \mathbb{F}^*_{p^m}\subseteq\mathbb{F}^*_{p^n}$ and ${\rm
gcd}(m-k,n)={\rm gcd}(m-k,2k)=d$ by Equality (1.2), again by Lemma
3, Equation (3.1) has either $0$, $1$, $2$, or $p^d+1$ roots in
$\mathbb{F}_{p^n}$. Thus, $g_{\delta,\gamma}(y)=0$  also  has either
$0$, $1$, $2$, or $p^d+1$ roots in $\mathbb{F}_{p^n}$.

\vspace{1mm}

(2) Suppose that
 $y_1$, $y_2$ are two different solutions of $g_{\delta,\gamma}(y)=0$. Then
$$
\begin{array}{rcl}y_1y_2(y_1-y_2)^{p^{m-k}}&=&y_1^{p^{m-k}+1}y_2-y_2^{p^{m-k}+1}y_1\\&=&\left(-\frac{(\gamma
y_1+\delta)y_2}{\delta^{p^{n-k}}}\right)-\left(-\frac{(\gamma
y_2+\delta)y_1}{\delta^{p^{n-k}}}\right)\\&
=&\frac{\delta(y_1-y_2)}{\delta^{p^{n-k}}},\end{array}$$ i.e.,
$$
y_1y_2=\delta^{1-p^{n-k}}(y_1-y_2)^{1-p^{m-k}}.$$ This together with
Equality (1.2) imply $$ \left(y_1y_2\right)^\frac{p^n-1}{p^d-1}=1.$$

(3) Suppose that $y_0$ is the unique solution of
 $g_{\delta,\gamma}(y)=0$. Since
$y=-\frac{\delta}{\gamma}x$, one has $x_0=-\frac{\gamma
y_0}{\delta}$ is the unique solution of Equation (3.1) in
$\mathbb{F}_{p^n}$. Since $c\in \mathbb{F}_{p^m}$, $x_0^{p^m}$ is
also a solution of Equation (3.1). As a consequence, one has
$x_0=x_0^{p^m}$, i.e., $x_0\in \mathbb{F}_{p^m}$.  Then, by Lemma 3,
one has
\begin{equation}\begin{array}{rcl}1&=&\left(x_0-1\right)^{\frac{p^m-1}{p^d-1}}\\&=&\left(-\frac{\gamma
y_0}{\delta}-1\right)^{\frac{p^m-1}{p^d-1}}\\&=&\left(-\frac{\gamma
y_0+\delta}{\delta}\right)^{\frac{p^m-1}{p^d-1}}\\&=&\left(\delta^{p^{n-k}-1}y_0^{p^{m-k}+1}\right)^{\frac{p^m-1}{p^d-1}}.\end{array}
\end{equation}
Similarly, if $y$ is a solution of $g_{\delta,\gamma}(y)=0$, one can
verify that $y^{p^m}\delta^{1-p^m}$ is also a solution of
$g_{\delta,\gamma}(y)=0$. Thus, $y_0=y_0^{p^m}\delta^{1-p^m}$, i.e.,
\begin{equation}y_0^{p^m-1}=\delta^{p^m-1}.
\end{equation}
Notice that $d\mid m$ and $d\mid k$ by Equality (1.2). Then,
Equalities (3.2) and (3.3) imply
$$\begin{array}{rcl}1&=&\delta^{^{\frac{(p^{n-k}-1)(p^m-1)}{p^d-1}}}y_0^{\frac{(p^{m-k}+1)(p^m-1)}{p^d-1}}\\
&=&y_0^{^{\frac{(p^{n-k}-1)(p^m-1)}{p^d-1}}}y_0^{\frac{(p^{m-k}+1)(p^m-1)}{p^d-1}}\\&=&y_0^{\frac{p^{m-k}}{p^d-1}(p^n-1)}.\end{array}
$$
This leads to
$$y_0^{\frac{p^n-1}{p^d-1}}=\left(y_0^{\frac{p^{m-k}}{p^d-1}(p^n-1)}\right)^{p^{m+k}}=1.$$
\hfill$\blacksquare$

\vspace{1mm}

{\it Remark 1:} (1) For given $\gamma$ and $\delta$ with
$\gamma\delta\not=0$, by Proposition 1(3), if $g_{\delta,\gamma}(y)$
has exactly one solution in $\mathbb{F}_{p^n}$, then the unique
solution is $(p^d-1)$-th power in $\mathbb{F}_{p^n}$.

\vspace{1mm}

(2) By Proposition 1(2), if $g_{\delta,\gamma}(y)$ has at least two
different solutions in $\mathbb{F}_{p^n}$, then all these solutions
are $(p^d-1)$-th powers in $\mathbb{F}_{p^n}$, or none of them are
$(p^d-1)$-th powers in $\mathbb{F}_{p^n}$. Over the finite field of
characteristic $2$, an analogy of Proposition 1(2) was obtained to
study the cross correlation between period-different $m$-sequences
(Proposition 2 of \cite{HZ08}). But the analogy of Proposition 1(3)
does not exist in the characteristic 2 case \cite{HZ08}.

\vspace{2mm}

With Proposition 1, the rank of the quadratic form
$\Pi_{\gamma,\delta}(x)$ can be determined as follows.

\vspace{1mm}

{\it Proposition 2:} For $\gamma\neq 0$ or $\delta\neq 0$, the rank
of $\Pi_{\gamma,\delta}(x)$ is either $n$, $n-d$, or $n-2d$.

\vspace{1mm}

{\it Proof: } The integer $p^{n-{\rm rank}(\Pi_{\gamma,\delta})}$ is
equal to the number of $z\in \mathbb{F}_{p^n}$ such that
\begin{equation}
\Pi_{\gamma,\delta}(x+z)=\Pi_{\gamma,\delta}(x)\end{equation} holds
for all $x\in \mathbb{F}_{p^n}$. Equation (3.4) holds if and only if
\begin{equation*}
Tr_{1}^{m}\left(\gamma(x+z)^{p^m+1}\right)+Tr_{1}^{n}\left(\delta(x+z)^{p^k+1}\right)=Tr_{1}^{m}(\gamma
x^{p^m+1})+Tr_{1}^{n}(\delta x^{p^k+1}),
\end{equation*}
or equivalently if and only if
\begin{equation*}
Tr_{1}^{m}\left(\gamma(xz^{p^m}+x^{p^m}z)\right)+Tr_{1}^{n}\left(\delta(xz^{p^k}+x^{p^k}z)\right)+Tr_{1}^{m}(\gamma
z^{p^m+1})+Tr_{1}^{n}(\delta z^{p^k+1})=0.
\end{equation*}
By the equality $xz^{p^m}+x^{p^m}z=Tr_m^n(xz^{p^m})$ and $\gamma \in
\mathbb{F}_{p^m}$, one has
\begin{equation*}
Tr_{1}^{n}(\gamma
xz^{p^m})+Tr_{1}^{n}\left(\delta(xz^{p^k}+x^{p^k}z)\right)+Tr_{1}^{m}(\gamma
z^{p^m+1})+Tr_{1}^{n}(\delta z^{p^k+1})=0,
\end{equation*}
i.e.,
\begin{equation*}Tr_{1}^{n}\left(\big(\gamma z^{p^m}+\delta z^{p^k}+(\delta
z)^{p^{n-k}}\big)x\right)+Tr_{1}^{m}(\gamma
z^{p^m+1})+Tr_{1}^{n}(\delta z^{p^{k}+1})=0.
\end{equation*}
Thus, Equation (3.4) holds for all $x\in \mathbb{F}_{p^n}$ if and
only if
\begin{equation} \gamma z^{p^m}+\delta z^{p^k}+(\delta
z)^{p^{n-k}}=0
\end{equation}
and
\begin{equation}
Tr_{1}^{m}(\gamma z^{p^m+1})+Tr_{1}^{n}(\delta z^{p^k+1})=0.
\end{equation}
By Equation (3.5), one has
\begin{equation*}\begin{array}{lll}Tr_{1}^{n}(\delta
z^{p^k+1})&=&-Tr_{1}^{n}(\gamma
z^{p^m+1}+\delta^{p^{n-k}}z^{p^{n-k}+1})\\&=&-Tr_{1}^{n}(\gamma
z^{p^m+1})-Tr^n_1(\delta^{p^{n-k}}z^{p^{n-k}+1})
\\&=&-2\,Tr_{1}^{m}(\gamma
z^{p^m+1})-Tr^n_1(\delta z^{p^k+1})
\end{array}\end{equation*}
since $\gamma z^{p^m+1}\in \mathbb{F}_{p^m}$, namely, Equation (3.5)
implies Equation (3.6). Then, it is sufficient to calculate the
number of solutions to Equation (3.5).

When $\delta=0$, one has $\gamma\neq 0$ and then Equation (3.5) has
one unique solution $z=0$. In the sequel, we only consider the case
 $\delta\neq 0$.

If $\gamma=0$, Equation (3.5) is equivalent to
$z(z^{p^{2k}-1}+\delta^{1-p^k})=0$. The number of all solutions to
this equation is $p^{{\rm gcd}(2k,n)}=p^{2d}$ or $1$ depending on
whether $-\delta^{1-p^k}$ is a $(p^{2d}-1)$-th power in
$\mathbb{F}_{p^n}$ or not. Thus, in this case the rank of
$\Pi_{\gamma,\delta}(x)$ is either $n-2d$ or $n$.

For $\gamma\neq 0$, $\gamma z^{p^m}+\delta z^{p^k}+(\delta
z)^{p^{n-k}}=z^{p^k}(\delta+\gamma
z^{p^m-p^k}+\delta^{p^{n-k}}z^{p^{n-k}-p^k})=0$. Thus, we only need
consider the number of nonzero solutions to the equation
\begin{equation} \delta+\gamma
z^{p^m-p^k}+\delta^{p^{n-k}}z^{p^{n-k}-p^k}=0.
\end{equation}
Let $y=z^{p^k(p^{m-k}-1)}$, then Equation (3.7) becomes
\begin{equation*} g_{\delta,\gamma}(y)=\delta^{p^{n-k}}y^{p^{m-k}+1}+\gamma
y+\delta=0.
\end{equation*}
By Proposition 1(1), $g_{\delta,\gamma}(y)=0$ has either $0$, $1$,
$2$, or $p^d+1$ roots in $\mathbb{F}_{p^n}$. Remark 1 and the fact
${\rm gcd}(p^k(p^{m-k}-1),p^n-1)=p^d-1$ show that Equation (3.7) has
$0$, $p^d-1$, $2(p^d-1)$, or $(p^d-1)(p^d+1)=p^{2d}-1$ nonzero
solutions in $\mathbb{F}_{p^n}$. Then, Equation (3.5) has $1$,
$p^d$, $2p^d-1$, or $p^{2d}$ solutions. Since $2p^d-1$ is not a
power of $p$ and hence is not the number of solutions of an
$F_p$-linearization polynomial, the number of solutions to Equation
(3.4) is equal to $1$, $p^d$, or $p^{2d}$.

\vspace{1mm}

Therefore, the rank of $\Pi_{\gamma,\delta}(x)$ is either $n$,
$n-d$, or $n-2d$. \hfill$\blacksquare$

\vspace{2mm}

In order to further determine the rank distribution of
$\Pi_{\gamma,\delta}(x)$, we define three sets
\begin{equation}
R_{i}={\Big\{}(\gamma,\delta)\mid {\rm
rank}(\Pi_{\gamma,\delta})=n-i, (\gamma,\delta)\in
\mathbb{F}_{p^m}\times \mathbb{F}_{p^n}\setminus \{(0,0)\}\,{\Big\}}
\end{equation}
for $i=0$, $d$, and $2d$. To achieve this goal, we need to evaluate
the exponential sum $S(\gamma,\delta,\epsilon)$ defined by
\begin{equation}
\begin{array}{lll}
S(\gamma,\delta,\epsilon)&=&\sum\limits_{x\in
\mathbb{F}_{p^n}}\omega^{\Pi_{\gamma,\delta}(x)+Tr_{1}^{n}(\epsilon
x)}
 \end{array}
\end{equation} where
 $\gamma\in\mathbb{F}_{p^m}, \delta, \epsilon\in \mathbb{F}_{p^n}$.
For $\rho\in \mathbb{F}_{p}$, let $N_{\gamma,\delta,\epsilon}(\rho)$
denote the number of solutions to the equation $\Pi_{\gamma,\delta
}(x)+Tr^n_1(\epsilon x)=\rho$. Then, the exponential sum can be
expressed as
\begin{equation}
S(\gamma,\delta,\epsilon)=\sum_{\rho=0}^{p-1}N_{\gamma,\delta,\epsilon}(\rho)\omega^{\rho}.
\end{equation}

\vspace{1mm}

Let $f(x)=\Pi_{\gamma,\delta}(x)$ be as in Equality (1.1). For
convenience, for $i\in \{0,d,2d\}$, we define
\begin{equation}\begin{array}{c}\Delta_i=(-1)^{\lfloor\frac{n-i}{2}\rfloor}
\prod\limits_{j=1}^{n-i}a_j,\end{array}\end{equation} where
$\lfloor\frac{n-i}{2}\rfloor$ denotes the largest integer not
exceeding $\frac{n-i}{2}$, and the coefficients $a_j$ are defined by
Equality (2.2).

\vspace{1mm}

In what follows, we will study the values of $S(\gamma,\delta,0)$
according to the rank of $\Pi_{\gamma,\delta}(x)$, and then use them
to determine the rank distribution.

\vspace{2mm}

{\it Case 1:} $(\gamma,\delta) \in R_0$.

\vspace{1mm}

In this case, ${\rm rank}(\Pi_{\gamma,\delta})=n$ and every
coefficient $a_i$ in Equality (2.2) is nonzero. Since ${\rm
det}(\Pi_{\gamma,\delta})(\det(B))^2=\prod\limits_{i=1}^{n}a_i$, one
has $\eta\left({\rm
det}(\Pi_{\gamma,\delta})\right)=\eta(\prod\limits_{i=1}^{n}a_i)$.
Then by Lemma 1,
$$\begin{array}{c}
 N_{\gamma,\delta,0}(\rho)=p^{n-1}+v(\rho)p^{\frac{n-2}{2}}\eta (\Delta_0)\end{array}
$$
and then by Equality (3.10),
$$\begin{array}{c}
S(\gamma,\delta,0)=\eta (\Delta_0)p^{\frac{n}{2}}\end{array}$$ since
$\sum\limits^{p-1}_{\rho=0}v(\rho)\omega^{\rho}=p$.

\vspace{1mm}

{\it Case 2:} $(\gamma,\delta)\in R_d$.

\vspace{1mm}

Since ${\rm rank}(\Pi_{\gamma,\delta})=n-d$, by Equality (2.2),
there exactly exist $d$ integers $i$ with $1\leq i\leq n$ such that
$a_i=0$. Without loss of generality, we assume
$\prod\limits_{i=1}^{n-d}a_{i}\neq 0$ and $a_i=0$ for $n-d<i\leq n$.
Then, $\Pi_{\gamma,\delta}(x)=\sum\limits_{i=1}^{n-d}a_{i}y_i^2,$
and by Lemma 1, for odd $d$,  one has
\begin{equation*}\begin{array}{lll}
 N_{\gamma,\delta,0}(\rho)&=&p^d\left(p^{n-d-1}+p^{\frac{n-d-1}{2}}\eta (\rho\Delta_d)\right)\\
 &=&p^{n-1}+p^{\frac{n+d-1}{2}}\eta (\rho\Delta_d).
\end{array}\end{equation*}
By Equality (3.10) and Lemma 2,
\begin{equation*}\begin{array}{lll}
S(\gamma,\delta,0)&=&\eta (\Delta_d)p^{\frac{n+d-1}{2}}\sum\limits_{\rho=0}^{p-1}\eta(\rho)\omega^\rho\\
 &=&\eta (\Delta_d)\sqrt{(-1)^{\frac{p-1}{2}}}p^{\frac{n+d}{2}}.
 \end{array}\end{equation*}

\vspace{2mm}

{\it Case 3:} $(\gamma,\delta)\in R_{2d}$.

\vspace{1mm}

Similarly as in Case 2, we can assume
$\prod\limits_{i=1}^{n-2d}a_{i}\neq 0$ and $a_i=0$ for $n-2d<i\leq
n$. Then, a similar analysis shows that
\begin{equation*}\begin{array}{c}
 N_{\gamma,\delta,0}(\rho)=p^{n-1}+v(\rho)p^{\frac{n+2d-2}{2}} \eta
 (\Delta_{2d})
\end{array}\end{equation*}
and
\begin{equation*}
\begin{array}{c}S(\gamma,\delta,0)=\eta (\Delta_{2d})p^{\frac{n}{2}+d}.\end{array}\end{equation*}

For each $i\in \{0, d, 2d\}$, we define two subsets of $R_i$ as
\begin{equation}\begin{array}{c}
R_{i,j}={\Big\{} (\gamma,\delta)\in R_i \mid \eta
(\Delta_{i})=j{\Big\}}
\end{array}\end{equation}
for $j=\pm 1$.

 \vspace{1mm}

Since $d$ is odd, the cardinality of $R_{d,1}$ can be proven to be
the same as that of $R_{d,-1}$ in the following lemma.

\vspace{1mm}

{\it Lemma 4:} $|R_{d,1}|=|R_{d,-1}|$.

 \vspace{1mm}

{\it Proof:}  Let $(\gamma,\delta)\in R_{d}$ and let $u\in
\mathbb{F}_{p}^*$ such that its inverse element satisfies
$\eta(u^{-1})=-1$. By Equality (3.9) and the analysis in Case 2, one
has
$$\begin{array}{lll}
S(u\gamma,u\delta,0)&=&\sum\limits_{x\in
\mathbb{F}_{p^n}}\omega^{\Pi_{u\gamma,u\delta}(x)}\\
&=&\sum\limits_{x\in
\mathbb{F}_{p^n}}\omega^{u\Pi_{\gamma,\delta}(x)}\\&=&\sum\limits_{\rho\in
\mathbb{F}_p}N_{\gamma,\delta,0}(u^{-1}\rho)w^\rho\\
&=&p^{\frac{n+d-1}{2}}\sum\limits_{\rho=0}^{p-1}\eta(u^{-1}\rho\Delta_d)\omega^\rho\\
&=&\eta(u^{-1})p^{\frac{n+d-1}{2}}\sum\limits_{\rho=0}^{p-1}\eta(\rho\Delta_d)\omega^\rho\\
&=&-S(\gamma,\delta,0).
 \end{array}$$
The above equality shows that for $j\in \{1,-1\}$, and if
$(\gamma,\delta)\in R_{d,j}$, then $(u\gamma,u\delta)\in R_{d,-j}$.
Thus, one has $|R_{d,1}|=|R_{d,-1}|$. \hfill$\blacksquare$

\vspace{2mm}

Applying Proposition 1 and Lemma 4, the cardinalities of $|R_d|$ and
$|R_{d,\pm 1}|$ can be determined as below.

\vspace{1mm}

{\it Proposition 3:} The set $R_d$ consists of $p^{m-d}(p^n-1)$
elements. Further, $|R_{d,\pm 1}|=p^{m-d}(p^n-1)/2$.

 \vspace{1mm}

 {\it Proof:} By Equality (3.8), it is sufficient to determine
 the number of $(\gamma,\delta)\in \mathbb{F}_{p^m}\times \mathbb{F}_{p^n}\setminus \{(0,0)\}$ such
 that Equation (3.5) has exactly $p^d$ solutions in
 $\mathbb{F}_{p^n}$. By the proof of Proposition 2, this case can
 occur only if $\gamma\delta\neq 0$.

 \vspace{1mm}

Let $W=\{x^{p^d-1}|\,\,x\in \mathbb{F}_{p^n}^*\}$ be the set of
nonzero $(p^d-1)$-th powers in $\mathbb{F}_{p^n}$. By Proposition
1(1), the equation $g_{\delta,\gamma}(y)=0$ in Proposition 2 has
either $0$, $1$, $2$, or $p^d+1$ roots in $\mathbb{F}_{p^n}$.

 \vspace{1mm}

When $g_{\delta,\gamma}(y)=0$ has at least two different roots in
$\mathbb{F}_{p^n}^*$, by Proposition 1(2) and Remark 1, if one of
the solutions belongs to $W$, then all these solutions are also in
$W$ and Equation (3.5) has at least $2(p^d-1)+1=2p^d-1$ solutions.
If none of these solutions belong to $W$, then Equation (3.5) has
only the unique solution $z=0$. Thus, in this case the number of
solutions to Equation (3.5) can never be $p^d$.

 \vspace{1mm}

If $g_{\delta,\gamma}(y)=0$ has exactly one solution in
$\mathbb{F}_{p^n}$, by Proposition 1(3) and Remark 1, Equation (3.5)
has $1\times(p^d-1)+1=p^d$ solutions. Since
$y=-\frac{\delta}{\gamma}x$, $g_{\delta,\gamma}(y)=0$ has exactly
one solution in $\mathbb{F}_{p^n}^*$ if and only if Equation (3.1)
has exactly one solution in $\mathbb{F}_{p^n}^*$. Furthermore, in
this case the unique solution to Equation (3.1) belongs to
$\mathbb{F}_{p^m}^*$ by the analysis in the proof Proposition 1(3).
When Equation (3.1) has exactly one solution in
$\mathbb{F}_{p^m}^*$, it has exactly one solution in
$\mathbb{F}_{p^n}^*$ since this equation has either $0$, $1$, $2$,
or $p^d+1$ roots in $\mathbb{F}_{p^n}^*$ and its solutions in
$\mathbb{F}_{p^n}\backslash \mathbb{F}_{p^m}$ occur in pairs.
Therefore, $g_{\delta,\gamma}(y)=0$ has exactly one solution in
$\mathbb{F}_{p^n}^*$ if and only if Equation (3.1) has exactly one
solution in $\mathbb{F}_{p^m}^*$.

 \vspace{1mm}

By Lemma 3, the number of $c\in \mathbb{F}_{p^m}^*$ such that
Equation (3.1) has exactly one solution in $\mathbb{F}_{p^m}$ is
$p^{m-d}$.  For any fixed $\gamma \in F_{p^m}^*$,
$c=\frac{\gamma^{p^{m-k}+1}}{\delta^{p^{m-k}(p^m+1)}}$ runs through
$\mathbb{F}_{p^m}^*$ exactly $(p^m+1)$ times when $\delta$ runs
through $\mathbb{F}_{p^n}^*$. Therefore, when $\gamma$ and $\delta$
run throughout $F_{p^m}^*$ and $F_{p^n}^*$, respectively, there are
exactly
\begin{equation*}|R_d|=(p^m-1)p^{m-d}(p^m+1)=p^{m-d}(p^n-1)
\end{equation*}
elements in $R_d$. This together with Lemma 4 finish the proof.
\hfill$\blacksquare$

\vspace{2mm}

The following proposition describes the sums of $i$-th powers of
$S(\gamma,\delta,0)$ for $1\leq i\leq 3$.

\vspace{1mm}

{\it Proposition 4:} (\textrm{i}) \begin{equation*}
 \sum\limits_{\gamma \in \mathbb{F}_{p^m}}\sum\limits_{\delta \in
 \mathbb{F}_{p^n}}S(\gamma,\delta,0)=p^{n+m}.
\end{equation*}

(\textrm{ii})\begin{equation*} \sum\limits_{\gamma \in
\mathbb{F}_{p^m}}\sum\limits_{\delta \in
\mathbb{F}_{p^n}}S(\gamma,\delta,0)^2=\left \{
 \begin{array}{cl}
p^{n+m},&p\equiv 3{\,\,\rm mod\,\,}4,\\
(2p^n-1)p^{n+m}, &p\equiv 1 {\,\,\rm mod\,\,}4.
\end{array}
 \right.
\end{equation*}

(\textrm{iii})\begin{equation*} \sum\limits_{\gamma \in
\mathbb{F}_{p^m}}\sum\limits_{\delta \in
\mathbb{F}_{p^n}}S(\gamma,\delta,0)^3=p^{n+m}(p^{n+d}+p^n-p^d).
\end{equation*}

 \vspace{1mm}

{\it Proof:} The proof of (\textrm{i}) is trivial, and we only give
the proof of (\textrm{ii}) and (\textrm{iii}).

(\textrm{ii}) It is true that
\begin{equation*}  \begin{array}{lll}&&\sum\limits_{\gamma \in
\mathbb{F}_{p^m},\delta \in \mathbb{F}_{p^n}}S(\gamma,\delta,0)^2\\
 &=&\sum\limits_{\gamma\in \mathbb{F}_{p^m},\delta \in \mathbb{F}_{p^n}}\sum\limits_{x,y \in
\mathbb{F}_{p^n}}\omega^{Tr_1^m \left(\gamma(x^{p^m+1}+y^{p^m+1})\right)+Tr_1^n \left(\delta(x^{p^k+1}+y^{p^k+1})\right)}\\
&=&\sum\limits_{x,y \in \mathbb{F}_{p^n}}\sum\limits_{\gamma\in
\mathbb{F}_{p^m}}\omega^{Tr_1^m\left(\gamma(x^{p^m+1}+y^{p^m+1})\right)}\sum\limits_{\delta
\in
\mathbb{F}_{p^n}}\omega^{Tr_1^n\left(\delta(x^{p^k+1}+y^{p^k+1})\right)}\\
&=&p^{n+m} |T_2|\\
\end{array}\end{equation*}
where $T_2$ consists of all solutions $(x,y)\in
\mathbb{F}_{p^n}\times \mathbb{F}_{p^n}$ to the following system of
equations
\begin{equation}
\left \{
 \begin{array}{c}
x^{p^m+1}+y^{p^m+1}=0,\\
x^{p^k+1}+y^{p^k+1}=0.\\
\end{array}
 \right.
\end{equation}

If $xy=0$, then $x=y=0$ by Equation (3.13).

If $xy\neq 0$, again by Equation (3.13), one has
\begin{equation*}(\frac{x}{y})^{p^k(p^{m-k}-1)}=1\end{equation*}
which implies
$(\frac{x}{y})^{p^{m-k}-1}=\left((\frac{x}{y})^{p^k(p^{m-k}-1)}\right)^{p^{n-k}}=1$,
that is to say, $\frac{x}{y}\in \mathbb{F}_{p^{m-k}}$ and then
\begin{equation*}\frac{x}{y}\in \mathbb{F}_{p^n}^* \cap \mathbb{F}_{p^{m-k}}^*=\mathbb{F}_{p^d}^* \end{equation*}
since ${\rm gcd}(m-k,n)={\rm gcd}(m-k,2k)=d$ by Equality (1.2). Let
$x=ty$ for some $t\in \mathbb{F}_{p^d}^*$, then  Equation (3.13)
becomes
\begin{equation}t^2+1=0\end{equation} since
$x^{p^m+1}=y^{p^m+1}(t^{p^{d\cdot \frac{m}{d}}})t=y^{p^m+1}t^2$ and
similarly $x^{p^k+1}=y^{p^k+1}t^2$.

\vspace{1mm}

For $p\equiv 3{\,\,\rm mod\,\,}4$, $-1$ is a non-square element in
$\mathbb{F}_{p^d}$ since $d$ is odd. Thus, Equation (3.14) has no
solutions and $|T_2|=1$. For $p\equiv 1{\,\,\rm mod\,\,}4$, $-1$ is
a square element in $\mathbb{F}_{p^d}$ and Equation (3.14) has 2
solutions in $\mathbb{F}_{p^d}^*$. Then, $|T_2|=1+2(p^n-1)=2p^n-1$.

\vspace{1mm} (\textrm{iii}) Similar analysis as in (\textrm{ii})
shows
\begin{equation*} \sum\limits_{\gamma \in
\mathbb{F}_{p^m},\delta \in \mathbb{F}_{p^n}}S(\gamma,\delta,0)^3
=p^{n+m} \mid T_3\mid,
\end{equation*}
where $T_3$ consists of all solutions $(x,y,z)\in
\mathbb{F}_{p^n}\times \mathbb{F}_{p^n}\times \mathbb{F}_{p^n}$ to
the following system of equations
\begin{equation}
\left \{
 \begin{array}{c}
x^{p^m+1}+y^{p^m+1}+z^{p^m+1}=0,\\
x^{p^k+1}+y^{p^k+1}+z^{p^k+1}=0.\\
\end{array}
 \right.
\end{equation}

For $xyz=0$, then $x=y=z=0$, or there are exactly two nonzero
elements in $\{x,y,z\}$. Thus, by Equation (3.13), in this case the
number of solutions to Equation (3.15) is equal to
$3(|T_2|-1)+1=3|T_2|-2$. For $xyz \neq 0$, the number of solutions
to Equation (3.15) is $(p^n-1)$ multiples of that to
\begin{equation} \left \{
 \begin{array}{c}
x^{p^m+1}+y^{p^m+1}+1=0,\\
x^{p^k+1}+y^{p^k+1}+1=0\\
\end{array}
 \right.
\end{equation}
where $x,y\in \mathbb{F}_{p^n}^*$. By Equation (3.16), one has
\begin{equation*}\left\{\begin{array}{l}
x^{(p^m+1)(p^{k}+1)}=\left(-(y^{p^{k}+1}+1)\right)^{p^m+1}=
y^{(p^m+1)(p^{k}+1)}+y^{(p^{k}+1)p^m}+y^{p^{k}+1}+1,
\\x^{(p^m+1)(p^{k}+1)}=\left(-(y^{p^{m}+1}+1)\right)^{p^k+1}=
y^{(p^k+1)(p^{m}+1)}+y^{(p^{m}+1)p^k}+y^{p^{m}+1}+1.
\end{array}\right.
\end{equation*}
This implies
\begin{equation*}
y^{(p^{k}+1)p^m}+y^{p^{k}+1}-y^{(p^{m}+1)p^k}-y^{p^{m}+1}=(y^{p^{m+k}}-y)(y^{p^m}-y^{p^k})=0.
\end{equation*}
By Equality (1.2), one has ${\rm gcd}(m+k,n)=d$. Then $y\in
\mathbb{F}_{p^{m+k}} \cap\, \mathbb{F}_{p^n}=\mathbb{F}_{p^d}$, or
$y\in \mathbb{F}_{p^{m}} \cap\, \mathbb{F}_{p^k}=\mathbb{F}_{p^d}$,
i.e., $y\in\mathbb{F}_{p^d}$.  Similarly, one has $x\in
\mathbb{F}_{p^d}$. By the fact ${\rm gcd}(m,k)=d$ and a similar
analysis as in the derivation of Equality (3.14), Equation (3.16) is
equivalent to
\begin{equation}
x^2+y^2+1=0,\,\, x,\, y \in \mathbb{F}^*_{p^d}.
\end{equation}
By Lemma 1, the number of all solutions $(x,y)\in
\mathbb{F}_{p^d}\times \mathbb{F}_{p^d}$ to the equation
$x^2+y^2+1=0$ is $p^d+v(-1)\eta(-1)$, where $\eta(-1)=1$ if $-1$ is
a square element in $F_{p^d}$, and $-1$ otherwise. Notice that the
solutions satisfying $xy=0$ and $x^2+y^2+1=0$ in $\mathbb{F}_{p^d}$
do not exist for $p\equiv 3\,\,{\rm mod}\,\,4$, and they are exactly
$(0,\pm \alpha^{\frac{p^n-1}{4}})$ and $(\pm
\alpha^{\frac{p^n-1}{4}},0)$ for $p\equiv 1\,\,{\rm mod}\,\,4$,
where $\alpha$ is a primitive element of $F_{p^n}$. As a
consequence, the number of all solutions to Equation (3.17) is given
by
\begin{equation*}
\left \{
 \begin{array}{cc}
p^d+1, & p\equiv 3\,\,{\rm mod}\,\,4,\\
(p^d-1)-4, & p\equiv 1\,\,{\rm mod}\,\,4.\\
\end{array}
 \right.
\end{equation*}
Thus,  one has
\begin{equation*}
\mid T_3\mid=1+(p^d+1)(p^n-1)=p^{n+d}+p^n-p^d
\end{equation*}
for $p\equiv 3\,\,{\rm mod}\,\,4,$ and \begin{equation*} \mid
T_3\mid=3(2p^n-1)-2+(p^d-5)(p^n-1)=p^{n+d}+p^n-p^d
\end{equation*}
for $p\equiv 1\,\,{\rm mod}\,\,4$.

\vspace{1mm}

Therefore, $|T_3|=p^{n+d}+p^n-p^d$. \hfill$\blacksquare$

\vspace{2mm}

With the above preparations, the rank distribution of
$\Pi_{\delta,\gamma}(x)$ can be determined as follows.

Since $S(0,0,0)=p^n$, by Proposition 4 and the values of
$S(\gamma,\delta,0)$ corresponding to the rank of
$\Pi_{\gamma,\delta}(x)$, one has
\begin{equation*}
\left \{
 \begin{array}{l}
p^{\frac{n}{2}}(|R_{0,1}|-|R_{0,-1}|)+p^{\frac{n}{2}+d}(|R_{2d,1}|-|R_{2d,-1}|)+p^n=\sum\limits_{\gamma
\in \mathbb{F}_{p^m}}\sum\limits_{\delta \in
 \mathbb{F}_{p^n}}S(\gamma,\delta,0),\\
p^n(|R_{0,1}|+|R_{0,-1}|)+(-1)^{\frac{p-1}{2}}p^{n+d}|R_d|+p^{n+2d}(|R_{2d,1}|+|R_{2d,-1}|)+p^{2n}\\\hspace{9cm}=\sum\limits_{\gamma
\in \mathbb{F}_{p^m}}\sum\limits_{\delta \in
 \mathbb{F}_{p^n}}S(\gamma,\delta,0)^2,\\
p^{\frac{3n}{2}}(|R_{0,1}|-|R_{0,-1}|)+p^{\frac{3n}{2}+3d}(|R_{2d,1}|-|R_{2d,-1}|)+p^{3n}=\sum\limits_{\gamma
\in \mathbb{F}_{p^m}}\sum\limits_{\delta \in
 \mathbb{F}_{p^n}}S(\gamma,\delta,0)^3.
\end{array}
 \right.
\end{equation*}
This together with the equality
$$|R_{0,1}|+|R_{0,-1}|+|R_{d}|+|R_{2d,1}|+|R_{2d,-1}|=p^{n+m}-1$$
as well as Proposition 3 gives
\begin{equation}
\left \{
 \begin{array}{lll}
|R_{0,1}|=\frac{p^d(p^m+1)(p^n-1)}{2(p^d+1)},\\
|R_{0,-1}|=\frac{(p^{n+d}-2p^n+p^d)(p^m-1)}{2(p^d-1)},\\
|R_{2d,1}|=0,\\
|R_{2d,-1}|=\frac{(p^{m-d}-1)(p^n-1)}{p^{2d}-1}.
\end{array}
 \right.
\end{equation}

Therefore, we have the following result.

\vspace{1mm}

{\it Proposition 5:} When $(\gamma,\delta)$ runs through
$F_{p^m}\times F_{p^n}\setminus\{(0,0)\}$, the rank distribution of
the quadratic form $\Pi_{\gamma,\delta}(x)$ is given by Table 2.
\begin{table}\caption{Rank distribution of the quadratic form $\Pi_{\gamma,\delta}(x)$}
\begin{center}
\begin{tabular}{|c|c|}
\hline Rank  &  Frequency\\
\hline $n$& $(p^{m+n+2d}+p^n+p^{m+d}-p^{m+n}-p^{m+n+d}-p^{2d})/(p^{2d}-1)$\\
\hline $n-d$& $p^{m-d}(p^n-1)$\\
\hline $n-2d$& $(p^{m-d}-1)(p^n-1)/(p^{2d}-1)$\\
 \hline
\end{tabular}
\end{center}
\end{table}

\vspace{2mm}

\section{ Weight Distribution of The Nonbinary Kasami Codes}

This section determines the weight distribution of the nonbinary
Kasami codes $\mathcal{C}^k$. Furthermore, we also give the
distribution of $S(\gamma,\delta,\epsilon)$, which will be used to
derive the correlation distribution of the sequence families
proposed in next section.

\vspace{1mm}

Since the weight of the codeword $c(\gamma,\delta,\epsilon)$ is
equal to
$p^n-1-(N_{\gamma,\delta,\epsilon}(0)-1)=p^n-N_{\gamma,\delta,\epsilon}(0)$,
it is sufficient to find $N_{\gamma,\delta,\epsilon}(0)$ for any
given $\gamma,\delta,\epsilon$.

\vspace{1mm}

Under the basis $\{\alpha_{1}, \alpha_{2},\cdots, \alpha_{n}\}$ of
$\mathbb{F}_{p^n}$ over $\mathbb{F}_{p}$, let
$\epsilon=\sum\limits_{i=1}^{n} \epsilon_i \alpha_i$ with
$\epsilon_i\in \mathbb{F}_{p}$. Then, $ Tr_1^n(\epsilon
x)=\Lambda^{{\rm T}}CX $ where $\Lambda^{{\rm
T}}=(\epsilon_{1},\epsilon_{2},\cdots,
\epsilon_{n})\in\mathbb{F}_{p}^n$ and the matrix
$C=\left(Tr_1^n(\alpha_i\alpha_j)\right)_{1\leq i,j\leq n}$, which
is nonsingular since $\{\alpha_{1},\alpha_{2},\cdots, \alpha_{n}\}$
is a basis of $\mathbb{F}_{p^n}$ over $\mathbb{F}_{p}$. Making a
nonsingular substitution $X=BY$ as in Section 2, one has
\begin{equation}\begin{array}{c}
\Pi_{\gamma,\delta}(x) +Tr_1^n(\epsilon x)=Y^{{\rm T}}B^{{\rm
T}}ABY+\Lambda^{{\rm T}}CBY
=\sum\limits_{i=1}^{n}a_{i}y_i^2+\sum\limits_{i=1}^{n}b_{i}y_i\end{array}
\end{equation}
where $\Lambda^{{\rm T}}CB=(b_{1}, b_{2},\cdots, b_{n})$. Then, for
any $\rho\in \mathbb{F}_{p}$, $\Pi_{\gamma,\delta}(x)
+Tr_1^n(\epsilon x)=\rho$ if and only if
$\sum\limits_{i=1}^{n}a_{i}y_i^2+\sum\limits_{i=1}^{n}b_{i}y_i=\rho.
$

\vspace{2mm}

We calculate the values of $N_{\gamma,\delta,\epsilon}(\rho)$
($\rho\in \mathbb{F}_{p}$) and the exponential sum
$S(\gamma,\delta,\epsilon)$ as follows:

\vspace{1mm}

{\it Case 1:} $(\gamma,\delta)=(0,0)$.

\vspace{1mm}

For $\epsilon\neq 0$, since the function $Tr_1^n(\epsilon x)$ is
linear from $\mathbb{F}_{p^n}$ to $\mathbb{F}_{p}$,  the weight of
$c(\gamma,\delta,\epsilon)$ is $(p-1)p^{n-1}$, and then
\begin{equation*} S(0,0,\epsilon)=\left \{
\begin{array}{ll}
0,\,\, \,\,\epsilon\neq 0,\\
p^n,\, \epsilon=0.\\
\end{array} \right.
\end{equation*}

\vspace{1mm}

{\it Case 2:} $(\gamma,\delta)\not=(0,0)$.

\vspace{1mm}

{\it Case 2.1:} $(\gamma,\delta) \in R_0$.

\vspace{1mm}

A substitution $y_i=z_i-\frac{b_i}{2a_i}$ for $1\leq i\leq n$ leads
to
\begin{equation*}
 \sum\limits_{i=1}^{n}(a_{i}y_i^2+b_{i}y_i)=\rho
 \Longleftrightarrow
 \sum\limits_{i=1}^{n}a_{i}z_i^2=\lambda_{\gamma,\delta,\epsilon}+\rho,
\end{equation*}
where  $\lambda_{\gamma,\delta,\epsilon}=
\sum\limits_{i=1}^{n}\frac{b_i^2}{4a_i}$. Then, for any $\rho\in
\mathbb{F}_p$ and given $(\gamma,\delta)\in R_0$, by Lemma 1, one
has
\begin{equation}\begin{array}{c}
 N_{\gamma,\delta,\epsilon}(\rho)=p^{n-1}+v(\lambda_{\gamma,\delta,\epsilon}+\rho)p^{\frac{n-2}{2}}\eta (\Delta_0).\end{array}
\end{equation}

When $\epsilon$ runs through $\mathbb{F}_{p^n}$,
$(b_1,b_2,\cdots,b_n)$ runs through $\mathbb{F}_p^n$ since $CB$ is
nonsingular. Notice that $\lambda_{\gamma,\delta,\epsilon}$ is  a
quadratic form with $n$ variables $b_i$ for $1\leq i\leq n$. Then,
for any given $(\gamma,\delta)\in R_0$, by Lemma 1, when $\epsilon$
runs through $\mathbb{F}_{p^n}$, one has
\begin{equation}\begin{array}{c}
\lambda_{\gamma,\delta,\epsilon}=\sum\limits_{i=1}^{n}\frac{b_i^2}{4a_i}=\rho'\,{\rm\,\,
occurs \,\,}p^{n-1}+v(\rho')p^{\frac{n-2}{2}}\eta (\Delta_0)
{\rm\,\, times \,\,}\end{array}
\end{equation}for each $\rho'\in \mathbb{F}_p$ since
 $\eta \Big(
 \frac{1}{4^n\prod\limits_{i=1}^{n}a_{i}}\Big)=\eta \left(
 \prod\limits_{i=1}^{n}a_{i}\right)$. Thus, when $\epsilon$ runs through
$\mathbb{F}_{p^n}$, by Equalities (4.2) and (4.3),
\begin{equation*}\begin{array}{c}N_{\gamma,\delta,\epsilon}(0)=p^{n-1}+(p-1)p^{\frac{n-2}{2}}\eta
(\Delta_0)\end{array}\end{equation*}
 occurs $p^{n-1}+(p-1)p^{\frac{n-2}{2}}\eta (\Delta_0)$ times, and
 \begin{equation*}\begin{array}{c}N_{\gamma,\delta,\epsilon}(0)=p^{n-1}-p^{\frac{n-2}{2}}\eta
(\Delta_0)\end{array}\end{equation*}
  occurs $(p-1)\left(p^{n-1}-p^{\frac{n-2}{2}}\eta (\Delta_0)\right)$ times.

By Equalities (3.10) and (4.2), one has
\begin{equation*}
 \begin{array}{lllll}
 S(\gamma,\delta,\epsilon)
 &=&\sum\limits_{\rho\in\mathbb{F}_{p}}\left(p^{n-1}+v(\lambda_{\gamma,\delta,\epsilon}+\rho)p^{\frac{n-2}{2}}\eta (\Delta_0)\right)\omega^{\rho}\\
&=&\eta
(\Delta_0)p^{\frac{n-2}{2}}\sum\limits_{\rho\in\mathbb{F}_{p}}
v(\lambda_{\gamma,\delta,\epsilon}+\rho)\omega^{\rho}\\
&=&\eta
(\Delta_0)p^{\frac{n-2}{2}}\omega^{-\lambda_{\gamma,\delta,\epsilon}}\sum\limits_{\rho\in\mathbb{F}_{p}}
v(\lambda_{\gamma,\delta,\epsilon}+\rho)\omega^{\rho+\lambda_{\gamma,\delta,\epsilon}}\\
&=&\eta
(\Delta_0)p^{\frac{n}{2}}\omega^{-\lambda_{\gamma,\delta,\epsilon}},
 \end{array}
\end{equation*}
where the fourth equal sign holds since
$\sum\limits_{\rho\in\mathbb{F}_{p}}
v(\lambda_{\gamma,\delta,\epsilon}+\rho)\omega^{\rho+\lambda_{\gamma,\delta,\epsilon}}=p$.
Notice that
$v(-\lambda_{\gamma,\delta,\epsilon})=v(\lambda_{\gamma,\delta,\epsilon})$.
By Equality (4.3),
 for given $(\gamma,\delta)\in R_0$, when $\epsilon$ runs through
$\mathbb{F}_{p^n}$,
\begin{equation*}\begin{array}{c}
 S(\gamma,\delta,\epsilon)=\eta(\Delta_0)p^{\frac{n}{2}}\omega^{\rho}{\rm\,\,
occurs \,\,}p^{n-1}+v(\rho)p^{\frac{n-2}{2}}\eta (\Delta_0) {\rm\,\,
times }\end{array}
\end{equation*}for each $\rho\in \mathbb{F}_p$.

\vspace{2mm}

 {\it Case 2.2:} $(\gamma,\delta)\in R_d$.

\vspace{1mm}

In this case, one has
\begin{equation*}
\Pi_{\gamma,\delta}(x) +Tr_1^n(\epsilon
x)=\sum\limits_{i=1}^{n-d}a_{i}y_i^2+\sum\limits_{i=1}^{n}b_{i}y_i.
\end{equation*}

If there exists some $b_i\neq 0$ for $n-d<i\leq n$, then for any
$\rho\in\mathbb{F}_p$, $N_{\gamma,\delta,\epsilon}(\rho)=p^{n-1}$
and by Equality (3.10)$, S(\gamma,\delta,\epsilon)=0$. Further, for
given $(\gamma,\delta)\in R_d$, when $\epsilon$ runs through
$\mathbb{F}_{p^n}$, there are exactly $p^n-p^{n-d}$ choices for
$\epsilon$ such that there is at least one $b_i\neq 0$ with
$n-d<i\leq n$ since $CB$ is nonsingular.

\vspace{1mm}

If  $b_i=0$ for all $n-d<i\leq n$, a similar analysis as in Case 2.1
shows that
\begin{equation*}
 \sum\limits_{i=1}^{n-d}(a_{i}y_i^2+b_{i}y_i)=\rho
 \Longleftrightarrow
 \sum\limits_{i=1}^{n-d}a_{i}z_i^2=\lambda_{\gamma,\delta,\epsilon}+\rho,
\end{equation*}
where $\lambda_{\gamma,\delta,\epsilon}=
\sum\limits_{i=1}^{n-d}\frac{b_i^2}{4 a_i}$ and
$z_i=y_i+\frac{b_i}{2a_i}$ for $1\leq i\leq n-d$. Then, for any
$\rho\in \mathbb{F}_p$ and given $(\gamma,\delta)\in R_d$, by Lemma
1, one has
\begin{equation}\begin{array}{rcl}
 N_{\gamma,\delta,\epsilon}(\rho)&=&p^{d}\left(p^{n-d-1}+p^{\frac{n-d-1}{2}}\eta \left((\lambda_{\gamma,\delta,\epsilon}+\rho)\Delta_{d}\right)\right)\\
&=&p^{n-1}+p^{\frac{n+d-1}{2}}\eta
\left((\lambda_{\gamma,\delta,\epsilon}+\rho)\Delta_{d}\right)
\end{array}
\end{equation}
since $n-d$ is odd. For given $(\gamma,\delta)\in R_d$, by Lemma 1,
when $(b_1,b_2,\cdots,
 b_{n-d})$ runs through $\mathbb{F}_{p}^{n-d}$, one has
\begin{equation}\begin{array}{c}
\lambda_{\gamma,\delta,\epsilon}=\sum\limits_{i=1}^{n-d}\frac{b_i^2}{4a_i}=\rho'{\rm\,\,
occurs \,\,}p^{n-d-1}+\eta(\rho')p^{\frac{n-d-1}{2}}\eta(\Delta_d)
{\rm\,\, times \,\,}\end{array}
\end{equation}
for each $\rho'\in \mathbb{F}_p$. Since there are $\frac{p-1}{2}$
square and non-square elements in $\mathbb{F}_{p^*}$, respectively,
$\eta(\lambda_{\gamma,\delta,\epsilon})=0$ occurs $p^{n-d-1}$ times,
and $\pm 1$ occur $\frac{p-1}{2}\left(p^{n-d-1}\pm
p^{\frac{n-d-1}{2}}\eta (\Delta_d)\right)$ times, respectively.
Therefore, when $(b_1,b_2,\cdots,
 b_{n-d})$ runs through $\mathbb{F}_{p}^{n-d}$,
\begin{equation*}\begin{array}{c}N_{\gamma,\delta,\epsilon}(0)=p^{n-1}\,\,{\rm occurs}\,\,p^{n-d-1}\,\,{\rm times,}\end{array}\end{equation*}
 and
 \begin{equation*}\begin{array}{c}N_{\gamma,\delta,\epsilon}(0)=p^{n-1}\pm p^{\frac{n+d-1}{2}}\eta
(\Delta_{d})\end{array}\end{equation*}
 occurs $\frac{p-1}{2}\left(p^{n-d-1}\pm
p^{\frac{n-d-1}{2}}\eta (\Delta_d)\right)$ times.

By Equality (4.4), one has
\begin{equation*}
 \begin{array}{rcl}
 S(\gamma,\delta,\epsilon)
 &=&\sum\limits_{\rho\in\mathbb{F}_{p}}\left(p^{n-1}+p^{\frac{n+d-1}{2}}\eta (
 (\lambda_{\gamma,\delta,\epsilon}+\rho)\Delta_d)\right)\omega^{\rho}\\
&=&\eta(\Delta_d)p^{\frac{n+d-1}{2}}\sum\limits_{\rho\in\mathbb{F}_{p}}
\eta(\lambda_{\gamma,\delta,\epsilon}+\rho)\omega^{\rho}\\
&=&\eta(\Delta_d)p^{\frac{n+d-1}{2}}\omega^{-\lambda_{\gamma,\delta,\epsilon}}
\sum\limits_{\rho\in\mathbb{F}_{p}}
\eta(\lambda_{\gamma,\delta,\epsilon}+\rho)\omega^{\rho+\lambda_{\gamma,\delta,\epsilon}}\\
&=&\eta(\Delta_d)p^{\frac{n+d}{2}}\sqrt{(-1)^{\frac{p-1}{2}}}\omega^{-\lambda_{\gamma,\delta,\epsilon}},
 \end{array}
\end{equation*}
where the fourth equal sign holds due to Lemma 2.

\vspace{1mm}

By Equality (4.5),
 for given $(\gamma,\delta)\in R_d$, when $(b_1,b_2,\cdots,
 b_{n-d})$ runs through $\mathbb{F}_{p}^{n-d}$,
\begin{equation*}\begin{array}{c}
 S(\gamma,\delta,\epsilon)=\eta(\Delta_d)p^{\frac{n+d}{2}}\sqrt{(-1)^{\frac{p-1}{2}}}\omega^{\rho}{\rm\,\,
occurs \,\,}p^{n-d-1}+\eta(-\rho)p^{\frac{n-d-1}{2}}\eta (\Delta_d)
{\rm\,\, times }\end{array}
\end{equation*}for each $\rho\in \mathbb{F}_p$.

\vspace{2mm}

{\it Case 2.3:} $(\gamma,\delta)\in R_{2d}$.

\vspace{1mm}

In this case, one has
\begin{equation*} \Pi_{\gamma,\delta}(x)
+Tr_1^n(\epsilon
x)=\sum\limits_{i=1}^{n-2d}a_{i}y_i^2+\sum\limits_{i=1}^{n}b_{i}y_i.
\end{equation*}

Similarly as in Case 2.2, if there exists some $b_i\neq 0$ with
$n-2d<i\leq n$, then $N_{\gamma,\delta,\epsilon}(\rho)=p^{n-1}$ for
any $\rho\in \mathbb{F}_p$, and $S(\gamma,\delta,\epsilon)=0$.
Further, for given $(\gamma,\delta)\in R_{2d}$, when $\epsilon$ runs
through $\mathbb{F}_{p^n}$, there are $p^n-p^{n-2d}$ choices for
$\epsilon$ such that there is at least one $b_i\neq 0$ with $n-2d<
i\leq n$.

\vspace{1mm}

If $b_i=0$ for all  $n-2d< i\leq n$, a similar analysis shows that
 for any $\rho\in
\mathbb{F}_p$ and given $(\gamma,\delta)\in R_{2d}$, by Lemma 1, one
has
\begin{equation}\begin{array}{c}
 N_{\gamma,\delta,\epsilon}(\rho)=p^{n-1}+v(\lambda_{\gamma,\delta,
 \epsilon}+\rho)p^{\frac{n+2d-2}{2}}\eta (\Delta_{2d}),\end{array}
 \end{equation}
where $\lambda_{\gamma,\delta,\epsilon}=
\sum\limits_{i=1}^{n-2d}\frac{b_i^2}{4 a_i}$.
 For given $(\gamma,\delta)\in R_{2d}$, when $(b_1,b_2,\cdots,
 b_{n-2d})$ runs through $\mathbb{F}_{p}^{n-2d}$, by Lemma 1, one
 has
\begin{equation}\begin{array}{c}
\lambda_{\gamma,\delta,\epsilon}=\sum\limits_{i=1}^{n-2d}\frac{b_i^2}{4a_i}=\rho'\,{\rm\,\,
occurs \,\,}p^{n-2d-1}+v(\rho')p^{\frac{n-2d-2}{2}}\eta(\Delta_{2d})
{\rm\,\, times \,\,}\end{array}
\end{equation}
for each $\rho'\in \mathbb{F}_p$. Thus, when $(b_1,b_2,\cdots,
 b_{n-2d})$ runs through $\mathbb{F}_{p}^{n-2d}$,
\begin{equation*}\begin{array}{c}N_{\gamma,\delta,\epsilon}(0)=p^{n-1}+(p-1)p^{\frac{n+2d-2}{2}}\eta
(\Delta_{2d})\end{array}\end{equation*}
  occurs $p^{n-2d-1}+(p-1)p^{\frac{n-2d-2}{2}}\eta
(\Delta_{2d})$ times, and
\begin{equation*}\begin{array}{c}N_{\gamma,\delta,\epsilon}(0)=p^{n-1}-p^{\frac{n+2d-2}{2}}\eta
(\Delta_{2d})\end{array}\end{equation*}
  occurs $(p-1)\left(p^{n-2d-1}-p^{\frac{n-2d-2}{2}}\eta
(\Delta_{2d})\right)$ times. By Equality (4.6), one has
\begin{equation*}
 \begin{array}{c}
 S(\gamma,\delta,\epsilon)=\eta(\Delta_{2d})p^{\frac{n}{2}+d}
 \omega^{-\lambda_{\gamma,\delta,\epsilon}}.
 \end{array}
\end{equation*}

By Equality (4.7),
 for given $(\gamma,\delta)\in R_{2d}$, when $(b_1,b_2,\cdots,
 b_{n-2d})$ runs through $\mathbb{F}_{p}^{n-2d}$,
\begin{equation*}\begin{array}{c}
 S(\gamma,\delta,\epsilon)=\eta(\Delta_{2d})p^{\frac{n}{2}+d}\omega^{\rho}{\rm\,\,
occurs \,\,}p^{n-2d-1}+v(\rho)p^{\frac{n-2d-2}{2}}\eta (\Delta_{2d})
{\rm\,\, times }\end{array}
\end{equation*}for each $\rho\in \mathbb{F}_p$.

For $i\in \{0,d,2d\}$ and $j\in \{1,-1\}$, since $\eta(\Delta_i)=j$
for $(\gamma,\delta)\in R_{i,j}$, Theorem 1 can be proven by the
above analysis, Equality (3.18), and Proposition 3.

\vspace{1mm}

{\it Proof of Theorem 1:} We only give the frequencies of the
codewords with weights $(p-1)p^{n-1}$,
$(p-1)(p^{n-1}-p^{\frac{n-2}{2}})$, and
$(p-1)p^{n-1}+p^{\frac{n+d-1}{2}}$. The other cases can be proven in
a similar way.

The weight of $c(\gamma,\delta,\epsilon)$ is equal to $(p-1)p^{n-1}$
if and only if $N_{\gamma,\delta,\epsilon}(0)=p^{n-1}$, which occurs
only in Cases 1, 2.2 and 2.3. The frequency is equal to
$$\begin{array}{rcl}&&p^n-1+\left((p^n-p^{n-d})+p^{n-d-1}\right)|R_d|+(p^n-p^{n-2d})|R_{2d}|\\
&=&(p^n-1)(1+p^{m+n-d}-p^{m+n-2d}+p^{m+n-2d-1}+p^{m+n-3d}-p^{n-2d}).\end{array}$$

The weight of $c(\gamma,\delta,\epsilon)$ is
$(p-1)(p^{n-1}-p^{\frac{n-2}{2}})$ if and only if
$N_{\gamma,\delta,\epsilon}(0)=p^{n-1}+(p-1)p^{\frac{n-2}{2}}$,
which occurs only in Case 2.1. The frequency is equal to
$$\begin{array}{c}(p^{n-1}+(p-1)p^{\frac{n-2}{2}})|R_{0,1}|=
p^d(p^m+1)(p^n-1)(p^{n-1}+(p-1)p^{\frac{n-2}{2}})/\left(2(p^d+1)\right).\end{array}$$

For $c(\gamma,\delta,\epsilon)$, its weight equals to
$(p-1)p^{n-1}+p^{\frac{n+d-1}{2}}$ if and only if
$N_{\gamma,\delta,\epsilon}(0)=p^{n-1}-p^{\frac{n+d-1}{2}}$, which
occurs only in Case 2.2. The frequency is equal to
$$\begin{array}{rcl}&&\frac{p-1}{2}(p^{n-d-1}+
p^{\frac{n-d-1}{2}}(-1))|R_{d,-1}|+\frac{p-1}{2}(p^{n-d-1}-
p^{\frac{n-d-1}{2}})|R_{d,1}|\\&=&
p^{m-d}(p^n-1)(p-1)(p^{n-d-1}-p^{\frac{n-d-1}{2}})/2.\end{array}$$\hfill$\blacksquare$

{\it Remark 2:} (1) From Table 1, the code $\mathcal{C}^k$ has $9$
different weights for  $d=1$, and $10$ different  weights for $d>1$.

(2) The codewords with weight $(p-1)(p^{n-1}-p^{\frac{n+2d-2}{2}})$
or $(p-1)p^{n-1}+p^{\frac{n+2d-2}{2}}$ do not exist in
$\mathcal{C}^k$ since $|R_{2d,1}|=0$.

\vspace{1mm}

The following result can also be similarly proven and we omit its
proof here.

\vspace{1mm}

{\it Proposition 6:} For $n=2m\geq 4$, the exponential sum
$S(\gamma,\delta,\epsilon)$ defined in Equality (3.9) has the
distribution given in Table 3
 when
$(\gamma,\delta,\epsilon)$ runs through
$\mathbb{F}_{p^m}\times\mathbb{F}_{p^n}\times\mathbb{F}_{p^n}$.

\vspace{1mm}

\begin{table}\caption{Distribution of $S(\gamma,\delta,\epsilon)$ (with $\rho$ varying in
$\mathbb{F}_p$)}
\begin{center}
\begin{tabular}{|c|c|}
\hline $S(\gamma,\delta,\epsilon)$ &  Frequency \\
\hline $ p^n $& $\,\,1\,\,$\\
\hline $ 0 $& $\,\,(p^n-1)\left(1+p^{m+n-d}-p^{m+n-2d}+p^{m+n-3d}-p^{n-2d}\right)$\\
\hline $ p^{\frac{n}{2}}\omega^\rho $& $\,\,p^d(p^m+1)(p^n-1)\left(p^{n-1}+v(\rho)p^{\frac{n-2}{2}}\right)/\left(2(p^d+1)\right)$\\
\hline
$\,\,-p^{\frac{n}{2}}\omega^\rho \,\, $& $(p^{n+d}-2p^n+p^d)(p^m-1)\left(p^{n-1}-v(\rho)p^{\frac{n-2}{2}}\right)/\left(2(p^d-1)\right)$\\
\hline
$\,\, p^{\frac{n+d}{2}}\sqrt{(-1)^{\frac{p-1}{2}}}\omega^\rho \,\, $& $p^{m-d}(p^n-1)(p^{n-d-1}+\eta(-\rho)p^{\frac{n-d-1}{2}})/2$\\
\hline
$\,\,-p^{\frac{n+d}{2}}\sqrt{(-1)^{\frac{p-1}{2}}}\omega^\rho \,\, $& $p^{m-d}(p^n-1)(p^{n-d-1}-\eta(-\rho)p^{\frac{n-d-1}{2}})/2$\\
 \hline
$\,\,-p^{\frac{n}{2}+d}\omega^\rho \,\, $& $(p^{m-d}-1)(p^n-1)\left(p^{n-2d-1}-v(\rho)p^{\frac{n-2d-2}{2}}\right)/(p^{2d}-1)$\\
 \hline
\end{tabular}
\end{center}
\end{table}

\vspace{2mm}

\section{A Class of Nonbinary Sequence Families}

By choosing cyclicly inequivalent codewords from $\mathcal{C}^k$, a
family of nonbinary sequences is defined by
\begin{equation}
\mathcal{F}^k={\Big\{}\{s_{a,b}(\alpha^t)\}_{0\leq t\leq
p^n-2}\,|\,\,a\in \mathbb{F}_{p^m}, b \in \mathbb{F}_{p^n}{\Big\}},
\end{equation}
where $\alpha$ is a primitive element of $\mathbb{F}_{p^n}$, and
\begin{equation*}
s_{a,b}(\alpha^t)=Tr_1^m(a\alpha^{(p^m+1)t})+Tr_1^n(b\alpha^{(p^{k}+1)t}+\alpha^{t}).\end{equation*}

The subfamily consisting of the sequences
$\{s_{a,0}(\alpha^t)\}_{0\leq t\leq p^n-2}$ for all $a\in
\mathbb{F}_{p^m}$ (and with fixing $b=0$) has been considered in
\cite{LK} and Appendix A of \cite{KM} respectively, and its
correlation distribution was determined in \cite{LK}. For $k=m+1$,
possible correlation values of the family $\mathcal{F}^{m+1}$ was
discussed in \cite{XZH}, but the correlation distribution remains
unsolved.

\vspace{1mm}

To aim at the correlation distribution of $\mathcal{F}^k$, we write
the correlation function between two sequences $s_{a_1,b_1}$ and
$s_{a_2,b_2}$ as
\begin{equation*}\begin{array}{lll}&&
C_{a_1b_1,a_2b_2}(\tau)\\&=&\sum\limits_{t=0}^{p^n-2}\omega^{s_{a_1,b_1}(t)-s_{a_2,b_2}(t+\tau)}\\
&=&\sum\limits_{t=0}^{p^n-2}\omega^{Tr_1^m\left((a_1-a_2\alpha^{(p^m+1)\tau})\alpha^{(p^m+1)t}\right)
+Tr_1^n\left((b_1-b_2\alpha^{(p^{k}+1)\tau})\alpha^{(p^{k}+1)t}+(1-\alpha^{\tau})\alpha^{t}\right)}\\
&=&-1+\sum\limits_{x\in
\mathbb{F}_{p^n}}\omega^{Tr_1^m\left((a_1-a_2\alpha^{(p^m+1)\tau})x^{p^m+1}\right)
+Tr_1^n\left((b_1-b_2\alpha^{(p^{k}+1)\tau})x^{p^{k}+1}+(1-\alpha^{\tau})x\right)}\\
&=&-1+S(\lambda_1,\lambda_2,\lambda_3),
\end{array}
\end{equation*}
where
\begin{equation}
\lambda_1=a_1-a_2\alpha^{(p^m+1)\tau},\,\,\lambda_2=b_1-b_2\alpha^{(p^{k}+1)\tau},\,\,
\lambda_3=1-\alpha^{\tau}.
\end{equation}

With this, the distribution of correlation values of the family
$\mathcal{F}^k$ can be described in terms of the exponential sum
$S(\lambda_1,\lambda_2,\lambda_3)$.

\vspace{1mm}

A simple property of $S(\gamma,\delta,\epsilon)$ is described as
below.

\vspace{1mm}

 {\it Lemma 5:
}For any given $\epsilon\in \mathbb{F}_{p^n}^*$, when
$(\gamma,\delta)$ runs through $
\mathbb{F}_{p^m}\times\mathbb{F}_{p^n}$, the distribution of
$S(\gamma,\delta,\epsilon)$  is the same as that of
$S(\gamma,\delta,1)$.

\vspace{1mm}

{\it Proof: } For any fixed $\epsilon\in \mathbb{F}_{p^n}^*$, one
has

\begin{equation*}\begin{array}{lll}
S(\gamma,\delta,\epsilon)&=&\sum\limits_{x\in
\mathbb{F}_{p^n}}\omega^{Tr_1^m(\gamma x^{p^m+1})
+Tr_1^n(\delta x^{p^{k}+1}+\epsilon x)}\\
&=&\sum\limits_{y\in
\mathbb{F}_{p^n}}\omega^{Tr_1^m(\gamma\epsilon^{-(p^m+1)}
y^{p^m+1})+Tr_1^n(\delta \epsilon^{-(p^{k}+1)}y^{p^{k}+1}+y)}\\
&=&S(\gamma \epsilon^{-(p^m+1)},\delta \epsilon^{-(p^{k}+1)},1).\\
\end{array}
\end{equation*}For any fixed $\epsilon\in \mathbb{F}_{p^n}^*$, when $(\gamma,\delta)$
runs through $ \mathbb{F}_{p^m}\times\mathbb{F}_{p^n}$, so does
$(\gamma \epsilon^{-(p^m+1)},\delta \epsilon^{-(p^{k}+1)})$. Thus,
the distribution of $S(\gamma,\delta,\epsilon)$  is the same as that
of $S(\gamma,\delta,1)$. \hfill$\blacksquare$

\vspace{1mm}

\begin{table*}\caption{Correlation
distribution of the sequence family $\mathcal{F}^k$ (with $\rho$
varying in $\mathbb{F}_{p}^*$)}
\begin{center}
\begin{tabular}{|c|c|}
\hline Correlation value & Frequency\\
\hline $p^n-1$& $p^{\frac{3n}{2}}$\\
\hline
$-1$& $p^{\frac{3n}{2}}(p^n-2)\left(1+p^{\frac{3n}{2}-d}-p^{\frac{3n}{2}-2d}+p^{\frac{3n}{2}-3d}-p^{n-2d}\right)$\\
 \hline
$p^{\frac{n}{2}}-1$& $p^{\frac{3n}{2}+d}(p^{\frac{n}{2}}+1)\left((p^n-2)(p^{n-1}+p^{\frac{n}{2}}-p^{\frac{n-2}{2}})+1\right)/\left(2(p^d+1)\right)$\\
\hline
$ p^{\frac{n}{2}}\omega^\rho-1$& $p^{\frac{3n}{2}+d}(p^{\frac{n}{2}}+1)(p^n-2)(p^{n-1}-p^{\frac{n-2}{2}})/\left(2(p^d+1)\right)$\\
\hline
$-p^{\frac{n}{2}}-1$& $p^{\frac{3n}{2}}(p^{n+d}-2p^n+p^d)\left((p^n-2)(p^{n-1}-p^{\frac{n}{2}}+p^{\frac{n-2}{2}})+1\right)/\left(2(p^{\frac{n}{2}}+1)(p^d-1)\right)$\\
 \hline
$-p^{\frac{n}{2}}\omega^\rho-1$&$p^{2n-1}(p^{n+d}-2p^n+p^d)(p^{n}-2)/\left(2(p^d-1)\right)$\\
\hline
$\pm p^{\frac{n+d}{2}}\sqrt{(-1)^{\frac{p-1}{2}}}-1$& $p^{2n-d}\left((p^{n}-2)p^{n-d-1}+1\right)/2$\\
\hline
$\begin{array}{c}p^{\frac{n+d}{2}}\sqrt{(-1)^{\frac{p-1}{2}}}\omega^\rho-1\end{array}$& $p^{2n-d}(p^{n}-2)\left(p^{n-d-1}+\eta(-\rho)p^{\frac{n-d-1}{2}}\right)/2$\\
\hline
$\begin{array}{c}- p^{\frac{n+d}{2}}\sqrt{(-1)^{\frac{p-1}{2}}}\omega^\rho-1\end{array}$& $p^{2n-d}(p^{n}-2)\left(p^{n-d-1}-\eta(-\rho)p^{\frac{n-d-1}{2}}\right)/2$\\
\hline
$-p^{\frac{n}{2}+d}-1$& $p^{\frac{3n}{2}}(p^{\frac{n}{2}-d}-1)\left((p^{n}-2)(p^{n-2d-1}-p^{\frac{n}{2}-d}+p^{\frac{n}{2}-d-1})+1\right)/(p^{2d}-1)$\\
\hline
$-p^{\frac{n}{2}+d}\omega^\rho-1$& $p^{\frac{3n}{2}}(p^{\frac{n}{2}-d}-1)(p^{n}-2)(p^{n-2d-1}+p^{\frac{n}{2}-d-1})/(p^{2d}-1)$\\
\hline
\end{tabular}
\end{center}
\end{table*}

 {\it Theorem 2:} Let $\mathcal{F}^k$ be the family of
sequences defined in Equality (5.1). Then, $\mathcal{F}^k$ is a
family of $p^{\frac{3n}{2}}$ nonbinary sequences with period
$p^n-1$, and its maximum correlation magnitude is equal to
$p^{\frac{n}{2}+d}+1$. Further, its correlation distribution is
given in Table 4.

\vspace{1mm}

{\it Proof: } By Equality (5.2), for any fixed $(a_2,b_2)\in
\mathbb{F}_{p^m}\times\mathbb{F}_{p^n}$, when $(a_1,b_1)$ runs
through $ \mathbb{F}_{p^m}\times\mathbb{F}_{p^n}$ and $\tau$ varies
from $0$ to $p^n-2$, $(\lambda_1,\lambda_2,\lambda_3)$ runs through
$\mathbb{F}_{p^m}\times\mathbb{F}_{p^n}\times\{\mathbb{F}_{p^n}\setminus
\{1\}\}$ one time. Thus, the correlation distribution of
$\mathcal{F}^k$ is $p^{\frac{3n}{2}}$ times as that of
$S(\gamma,\delta,\epsilon)-1$ when $(\gamma,\delta,\epsilon)$ runs
through
$\mathbb{F}_{p^m}\times\mathbb{F}_{p^n}\times\{\mathbb{F}_{p^n}\setminus
\{1\}\}$. By Proposition 3, Equality (3.18) and the possible values
of $S(\gamma,\delta,0)$ corresponding to $(\gamma,\delta)$, the
distribution of $S(\gamma,\delta,0)-1$ is obtained when
$(\gamma,\delta)$ runs through $
\mathbb{F}_{p^m}\times\mathbb{F}_{p^n}$. This together with
Proposition 6 give the distribution of $S(\gamma,\delta,\gamma)-1$
as $(\gamma,\delta,\epsilon)$ runs through
$\mathbb{F}_{p^m}\times\mathbb{F}_{p^n}\times\mathbb{F}^*_{p^n}$. By
Lemma 5, the distribution of $S(\gamma,\delta,\epsilon)-1$ can be
determined when $(\gamma,\delta,\epsilon)$ runs through
$\mathbb{F}_{p^m}\times\mathbb{F}_{p^n}\times\{\mathbb{F}^*_{p^n}\setminus
\{1\}\}$. Together with the distribution of $S(\gamma,\delta,0)-1$
as $(\gamma,\delta)$ runs through
$\mathbb{F}_{p^m}\times\mathbb{F}_{p^n}$, the correlation
distribution is given as Table 4.

By the definition of the sequence $\{s_{a,b}(\alpha^t)\}_{0\leq
t\leq p^n-2}$ , it is easy to check its period is $p^n-1$. From the
correlation distribution given in Table 4, one easily knows there
are exactly $p^{\frac{3n}{2}}$ sequences in the family
$\mathcal{F}^k$ and the maximum magnitude is $p^{\frac{n}{2}+d}+1$.

This finishes the proof.\hfill$\blacksquare$

{\it Remark 3:} (1) By Table 4, the correlation function of
$\mathcal{F}^k$ takes $5p+2$ values. Take $k=m-t$ for any odd
integers $t$ relatively prime to $m$ as stated in Section 2, then
$d=1$ and the families $\mathcal{F}^k$ have the maximum magnitude
$p^{\frac{n}{2}+1}+1$.

(2) Notice that if we remove the term $Tr_1^n(\alpha^{t})$ in
Equality (5.1) and define a sequence set as
$${\Big\{}
\{Tr_1^m(a\alpha^{(p^m+1)t})+Tr_1^n(b\alpha^{(p^{k}+1)t})\}_{0\leq
t\leq p^n-2}\mid a\in \mathbb{F}_{p^m},\,b\in
\mathbb{F}_{p^{n}}{\Big\}},$$ the period of each sequence in that
set is not larger than $(p^n-1)/2$. The sequence families proposed
in the present paper do not contain any sequences in that set.

\section{Concluding Remarks}

For an even integer $n=2m\geq 4$ and any positive integer $k$
satisfying Equality (1.2), a class of binary Kasami codes has been
extended to odd $p$-ary case. These $p$-ary $[p^n-1,{\frac{5n}{2}}]$
codes $\mathcal{C}^k$ can achieve the maximal value
$(p-1)p^{n-1}-p^{\frac{n}{2}}$ for their minimum distances by taking
$d=1$, and in this case, a family $\mathcal{F}^k$ of
$p^{\frac{3n}{2}}$ $p$-ary sequences with period $p^n-1$ can be
defined and have the maximum correlation magnitude
$p^{\frac{n}{2}+1}+1$.

We had tried to remove the restriction in Equality (1.2) on the
parameters $m$ and $k$. Clearly, the assumption of Equality (1.2) is
equivalent to say that $d={\rm gcd}(m,k)$ is odd and $m/d$ and $k/d$
have different parity. If $d$ is even, Proposition 2 still holds and
hence the rank of $\Pi_{\gamma,\delta}$ is still $n$, $n-d$, or
$n-2d$. If both $m/d$ and $k/d$ are odd, then the rank of
$\Pi_{\gamma,\delta}$ can be shown to be $n$, $n-2d$, or $n-4d$. In
each of these cases, Lemma 4 can not be proven by the method in this
paper since the ranks of the $\Pi_{\gamma,\delta}$ are all even, and
new equalities on the $|R_{n-{\rm rank}(\Pi_{\gamma,\delta}),\pm
1}|$ need to be established for determining the weight distribution.

With the help of a computer, for $p=3$ and $(m,k)=(3,1)$,  $(4,2)$,
the minimum distance of the $[p^n-1,{\frac{5n}{2}}]$ code
$\mathcal{C}^k$ is verified to be $(p-1)p^{n-1}-p^{\frac{n}{2}+1}$,
which is strictly less than $(p-1)p^{n-1}-p^{\frac{n}{2}}$.

It seems that new methods should be developed to deal with the
remaining case of $m$ and $k$.


\bibliographystyle{amsalpha}

   \end{document}